# SPEAKE(a)R: Turn Speakers to Microphones for Fun and Profit

(demo: https://www.youtube.com/watch?v=ez3o8aIZCDM)

Mordechai Guri, Yosef Solewicz, Andrey Daidakulov, Yuval Elovici

Ben-Gurion University of the Negev

Cyber Security Research Center

gurim@post.bgu.ac.il; yosef.solewicz@gmail.com; daidakul@post.bgu.ac.il; elovici@post.bgu.ac.il

*Abstract*

It's possible to manipulate the headphones (or earphones) connected to a computer, silently turning them into a pair of eavesdropping microphones – with software alone. The same is also true for some types of loudspeakers. This paper focuses on this threat in a cyber-security context. We present SPEAKE(a)R, a software that can covertly turn the headphones connected to a PC into a microphone. We present technical background and explain why most of today's PCs and laptops are susceptible to this type of attack. We examine an attack scenario in which malware can use a computer as an eavesdropping device, even when a microphone is not present, muted, taped[1], or turned off. We measure the signal quality and the effective distance, and survey the defensive countermeasures.

## 1. Introduction

Audio playing equipment such as loudspeakers, headphones, and earphones are widely used in PCs, laptops, smartphones, media entertainment systems, and more. In this section we refer to the any audio playing equipment that contains speakers (loudspeakers, headphones, earphones, etc.) as *speakers*.

Speakers aim at amplifying audio streams *out*, but in fact, a speaker can be seen as a microphone working in reverse mode: loudspeakers convert electric signals into a sound waveform, while microphones transform sounds into electric signals. Speakers use the changing magnetic field induced by electric signals to move a diaphragm in order to produce sounds. Similarly, in microphone devices, a small diaphragm moves through a magnetic field according to a sound's air pressure, inducing a corresponding electric signal [1]. This bidirectional mechanism facilitates the use of simple headphones as a feasible microphone, simply by plugging them into the PC microphone jack.

---

[1] "Why has Mark Zuckerberg taped over the webcam and microphone on his MacBook"? [32]

## 1.2 Speaker retasking

A typical computer chassis contains a number of audio jacks, either in the front panel, rear panel, or both. These jacks are the sockets for plugging in various audio equipment such as speakers, headphones, and microphones. Each jack is used either for input (line in), or for output (line out). The audio ports usually have a conventional coloring system; typically green is used for speakers (output jack), blue for line in (input jack), and pink for microphones (input jack).

Interestingly, the audio chipsets in modern motherboards and sound cards include an option to change the function of an audio port at a software level, a type of audio port programming sometimes referred to as *jack retasking* or *jack remapping*. This option is available on Realtek's (Realtek Semiconductor Corp.) audio chipsets, which are integrated into a wide range of PC motherboards today. Jack retasking – although documented in applicable technical specifications – is not well known, as was mentioned by the Linux audio developer, David Henningsson, in his blog:

*"Most of today's built-in sound cards are to some degree retaskable, which means that they can be used for more than one thing. …the kernel exposes an interface that makes it possible to retask your jacks, but almost no one seems to use it, or even know about it"* [2].

## 1.3 The threat

The fact that headphones and earphones are physically built like microphones, coupled with the fact that an audio port's role in the PC can be altered programmatically from output to input, creates a vulnerability which can be abused by hackers. A malware can stealthy reconfigure the headphone jack from a line out jack to a microphone jack. As a result, the connected headphones can function as a pair of recording microphones, thereby rendering the computer into an eavesdropping device – even when the computer doesn't have a connected microphone (Figure 1).

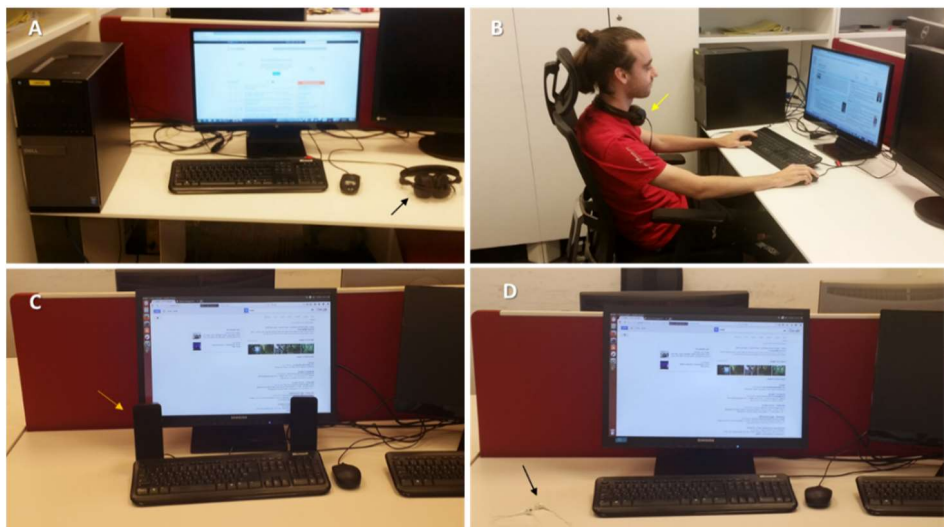

Figure 1. Illustration of SPEAKE(a)R. Headphones, speakers, and earphones are connected to a computer which have no microphone. A malware within the computer turn them into microphones. Note that in scenario C the method required 'passive' loudspeakers, which are rarely in use today.

In this paper we provide a technical overview of SPEAKE(a)R – a malware that can covertly transform headphones into a pair of microphones – and show how it can be exploited in different scenarios. We also evaluate its efficacy and the signal quality, and present defensive countermeasures.

The rest of this paper is structured as follows: Technical background is provided in section 2. Section 3 discusses threat scenarios. Section 4 describes the experimental results. Section 5 presents related work. Countermeasures are discussed in section 6. We conclude in Section 7.

## 2. Technical Background

The fact that speakers can be used in reverse mode, and hence, can function as microphones has been known for years, and is well documented in professional literature [1] [3]. However, this by itself doesn't raise a security concern, since it requires a speaker to be intentionally plugged into the microphone jack (an input port) in order to play the role of a microphone.

A glance into security related issues of such a 'speaker-as-microphone' scenario can be found in a partially declassified document released by the NSA in 2000, in response to an appeal of an earlier Freedom of Information Act (FOIA) request. The document is a guide to the installation of system equipment that takes into account red/black security concerns. It contains the following paragraph:

*"In addition to being a possible fortuitous conductor of TEMPEST emanations, **the speakers in paging, intercom and public address systems can act as microphones and retransmit classified audio discussions out of the controlled area via the signal line distribution.** This microphonic problem could also allow audio from higher classified areas to be heard from speakers in lesser classified areas. Ideally. Such systems should not be used. Where deemed vital, the following precautions should be taken in full or in part to lessen the risk of the system becoming an escape medium for NSI."* (NSTISSAM TEMPEST/2-95, RED/BLACK INSTALLATION [4] [5]).

### 2.1 Speakers, Headphones, and Earphones

Dynamic microphones are the inverse of dynamic loudspeakers. In the former, sound pressure variations move a membrane attached to a coil of wire in a magnetic field, generating an electric current/voltage. The inverse happens with loudspeakers: the electric voltage associated with a sound drives an electric current through a coil in the magnetic field, generating a force on the coil and moving the membrane attached to it, producing sound in the air (Figure 2). In fact, in simple intercom systems the same mechanism is used as either a microphone or loudspeaker.

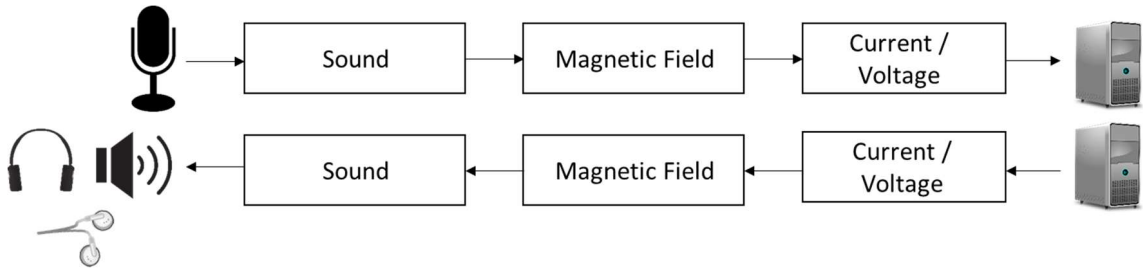

**Figure 2.** Basic illustration of audio recording and playing, demonstrating that the speaker and microphone are inverses of each other.

**Active Loudspeakers vs. Passive Loudspeakers**

Note, however, that the reversibility principle poses a limitation: the speaker must be passive (unpowered), without amplifier transitions. In the case of an active (self-powered) speaker, there is an amplifier between the jack and the speaker, hence the signal won't be passed from the output to the input side [6]. Since most modern loudspeakers have an internal amplifier [7], the threat presented in this paper is primarily relevant to headphones and earphones, and not to the loudspeakers typically connected to a PC.

## 2.2 Jack Retasking/Remapping

Desktop PCs may have a built-in (onboard) audio chip or external sound card. Typical PCs include various analog input and output jacks (audio ports). Input jacks are used for microphones or other sources of audio stream. The input data is sampled and processed by the audio hardware. Output jacks are used for loudspeakers, headphones, and other analog output playback devices. As noted, the capability of changing the functionality of the audio jacks is referred to as jack retasking or jack remapping.

Intel High Definition (HD) Audio (the successor of AC'97) is the standard audio component on modern PCs. Jack retasking is now part of the Intel High Definition Audio specification [8]. In this standard, the audio chip on the motherboard is referred to as an *audio codec.* The audio codec communicates with the host via a PCI or other system bus. Realtek Semiconductor Corp. provides the audio codec chip for many motherboards and chipset manufacturers, and is thus integrated in a wide range of desktop workstations and notebooks. The most common Realtek codecs for PCs are the multi-channel high definition audio codec series presented in Table 1. As noted in the table, the listed codec chips support jack retasking.

**Table 1. Realtek codec chips that support jack retasking**

| Realtek codec chips (all support jack retasking) | Integrated in | |
|---|---|---|
| ALC892, ALC889, ALC885, ALC888, ALC663, ALC662, ALC268, ALC262, ALC267, ALC272, ALC269, ALC3220 | Desktop PCs, Notebooks | 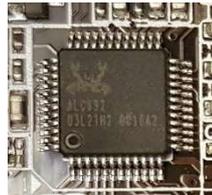 |

In this paper we primarily focus on Realtek codec chips, since they are the most common codecs in PCs. It's important to note that other codec manufacturers support jack retasking as well [9] [10].

2.2.1 Hardware Interface

At the hardware level, jack retasking means that the retaskable audio jacks are connected with both analog to ADC (analog-to-digital convertor) and DAC (digital-to-analog convertor) components, and hence can operate as input (microphone) or output (speaker) signal ports.

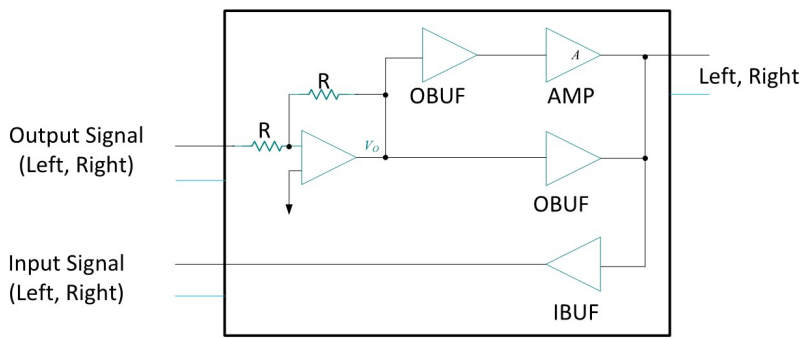

Figure 3. Jack retasking at the hardware level.

Figure 3 shows input/output retasking at the hardware level, based on the Realtek design [11]. The schematic diagram of the mic/line physical sockets illustrates two electrical circuits connected to the same physical socket. Choosing the input configuration enables the leg of the IBUF to rise, powering on the buffer, and allowing the signal to enter the computer. In contrast, choosing the output configuration enables the legs of the OBUF and the AMP to rise and enables the output buffer and amplifier. This action allows the signal to flow out from computer to the socket. When buffers are not enabled, no signal can go through.

2.2.1　HD Audio Codec Interface

The HD audio component consists of the controller and codec chips on the HD audio bus. Each codec contains various type of *widgets*, which are logical components that operate within the codec. Software can send messages (or *verbs*) to widgets in order to read or modify their settings. Such verbs are sent via the HDA link, which is a serial interface that connect the audio codec to the host PC. Typical messages to the audio codec are structured as `[NID][Verb][Payload]`, where `NID` is the node identifier (e.g., the widget to operate on), `Verb` is the type of operation (e.g., set configuration), and `Payload` contains the parameters for the operation (e.g., the configuration parameters).

The HD audio codec defines a number of *pin widgets*. Each pin, including the audio jack ports, has its own configuration. The configuration includes the jack color, location (rear, front, top, etc.), connection type (in or out), and other properties. For example, in the Realtek ALC892 chip, pins 14-17 (LINE2-L, LINE2-R, MIC2-L, MIC2-R), pins 21-24 (MIC1-L, MIC1-R, LINE1-L, LINE1-R), and pins 35-36 (FRONT-L, FRONT-R) are the analog input and output pins. In the retaskable pins (e.g., 14-17), it is possible to change the default configuration and its functionality, from *out* (e.g., headphone or speaker) to *in* (microphone), and vice versa. The HDA specification defines the complete codec architecture that allows a software driver to control various types of operations [8].

### 2.2.2 Operating System Interface

The vendors of audio codec chips, such as Realtek and Conexant, provide kernel drivers which implement the codec's functionality, including retasking, and expose it to the user mode programs. For example, the Realtek driver for Microsoft Windows allows remapping the audio jack via specific values in the Windows Registry. A guide for how to remap Realtek onboard jacks in Microsoft Windows can be found in [12]. The Linux kernel, a part of the Advanced Linux Sound Architecture (ALSA), exposes an interface that enables the jack configuration; the `hda-jack-retask` tool is a user mode program for Linux that allows the manipulation of the HD audio pins' control via a GUI interface [13].

## 3. Threat Scenarios

There are two main threat scenarios for using headphones as a microphone. The first scenario involves a PC that is not equipped with a microphone (or in which the microphone is muted or turned off) but has connected headphones, earphones, or passive speakers. In this scenario, a malware installed on the computer may reconfigure the headphone jack into a microphone jack. Usually (during normal system behavior), such reconfiguration takes place only while the headphones are not in use, such as when audio output is triggered (e.g., the user is playing music), and the microphone jack is instantly reconfigured back into a headphone jack. In the second scenario, the computer may be equipped with both microphone and headphones, but the headphones are better positioned for the desired recording, e.g., headphone are closer to the voice source and hence, can achieve better recording quality.

# 4. Experimental Results

The following experiments were conducted in order to assess the efficacy of using a headphone as a microphone as follows. Initially, we use headphones (instead of a microphone) to record environment sounds and investigate the effect of this on audio quality. Then we investigate the headphones' effectiveness as a receiver in digital communication protocols.

A series of audio quality measurements were collected in order to evaluate audio degradation when the audio is recorded via headphones instead of standard microphones. To measure speech intelligibility in different experimental setups, we used a set of pre-recorded sentences as a reference. More specifically, we selected a list of phonetically balanced sentences in English as our clean audio reference [14]. The list is comprised of simple phrases containing phonemes (in the same proportion as spoken English), which are often used for standardized development and testing of telecommunication systems, from cellphones to Voice over IP. This methodology enables quick and automatic evaluations of speech coding protocols.

The actual reference audio used during the experiments was taken from the open speech repository [15]. We used an 8 kHz recording of one of the lists by a male speaker. The audio was played at a high volume through commercial multimedia computer speakers (Genius SP-S110) and recorded using an off the shelf microphone device (Silverline MM202) and headphones (Sennheiser HD 25-1 II). Several objective speech quality measures were evaluated to estimate the degradation associated with the use of the headphones as described below. The experimental setups varied based on the distance between the computer playing the sound and the recording computer. In addition, we encoded and decoded the pre-recorded audio in order to simulate a situation in which the locally recorded audio is further transmitted from the recording computer to a remote computer.

**Objective Measures of Speech Quality**

The objective speech quality measures used in this research were estimated through the SNReval toolbox [16]. For each experimental setup we used the five objective speech quality measures provided by the toolbox and listed below. Implementation details for these measures, as well as additional information, can be found in [16].

1. NIST STNR
2. WADA SNR
3. SNR_VAD
4. BSS_EVAL (SAR)
5. PESQ

The first three measures focus on some version of signal-to-noise ratio (SNR), the ratio between the energy of some speech signal to that of its contaminating noise. In contrast, the last two measures reflect the distortion level of the recorded speech signal with respect to the reference pre-recorded (played aloud) signal and are more directly related to the intelligibility level of the recorded speech.

## 4.1 Results

**Table 2. Reference**

| NIST STNR (dB) | WADA SNR (dB) | SNR VAD (dB) |
|---|---|---|
| 40.5 | 49.2 | 10.1 |

**Table 3. Headphones**

| Distance (m) | NIST STNR (dB) | WADA SNR (dB) | SNR VAD (dB) | SAR (dB) | PESQ MOS |
|---|---|---|---|---|---|
| 1 | 7.5 | 3.0 | 2.8 | 3.7 | 2.6 |
| 3 | 7.0 | -20.0 | -7.2 | -2.8 | 2.0 |
| 5 | 6.5 | -20.0 | -6.1 | -5.9 | 2.0 |
| 7 | 7.8 | -2.4 | -11.0 | -5.5 | 2.2 |
| 9 | 8.0 | -10.4 | -20.6 | -4.3 | 2.0 |

**Table 4. Microphone**

| Distance (m) | NIST STNR (dB) | WADA SNR (dB) | SNR VAD (dB) | SAR (dB) | PESQ MOS |
|---|---|---|---|---|---|
| 1 | 29.0 | 22.6 | 6.7 | 8.7 | 2.5 |
| 9 | 13.8 | 8.0 | 4.8 | -3.8 | 2.0 |

**Table 5. ACC coding/decoding**

| Distance (m) | NIST STNR (dB) | WADA SNR (dB) | SNR VAD (dB) | SAR (dB) | PESQ MOS |
|---|---|---|---|---|---|
| 0 (Ref.) | 39.5 | 54.4 | 2.1 | 12.0 | 3.5 |
| 1 | 7.5 | 4.9 | -2.0 | 2.7 | 2.5 |
| 5 | 6.5 | -1.2 | -9.4 | -5.6 | 1.9 |
| 9 | 7.8 | -1.2 | -19.2 | -4.5 | 1.9 |

Table 2 provides the SNR measurements in decibels (dB) of the reference signal used in the experiments, namely, a sequential reading of sentences in the list (previously described), resulting in a file 40 seconds long. Note that SAR and PESQ measures are not estimated since this table addresses the pre-recorded reference signal alone.

Tables 3 and 4 (respectively) list the results for the quality measures for the headphone as microphone and microphone scenarios for different recording distances.

In Table 5 we show the quality degradation of the reference signal and headphone recorded speech after encoding and decoding, reproducing a subsequent transmission of the acquired speech via the Internet. The codec used was the Advanced Audio Coding (AAC) [17], the potential MP3 successor and the default codec for YouTube, the iPhone, iPod, and other media devices. This codec has a compression ratio of approximately 10 to one.

We note that, in general, SNR measurements are highly dependent on an accurate segmentation of speech versus noise excerpts. Therefore, in order to optimize segmentation accuracy and consistency across the different setups investigated, voice activity detection (VAD) was estimated for the reference and not recorded signals. This segmentation was then applied to reference and recorded pairs of signals after they have been time-aligned through cross-correlation.

**Frequency Domain**

In addition to the several objective SNR measures presented, we also provide frequency domain graphs corresponding to features used in the above calculations, comparing four transmission setups: (1) microphone, one meter apart (2) headphones, one meter apart, (3) headphones, five meters apart, and (4) headphones, nine meters apart.

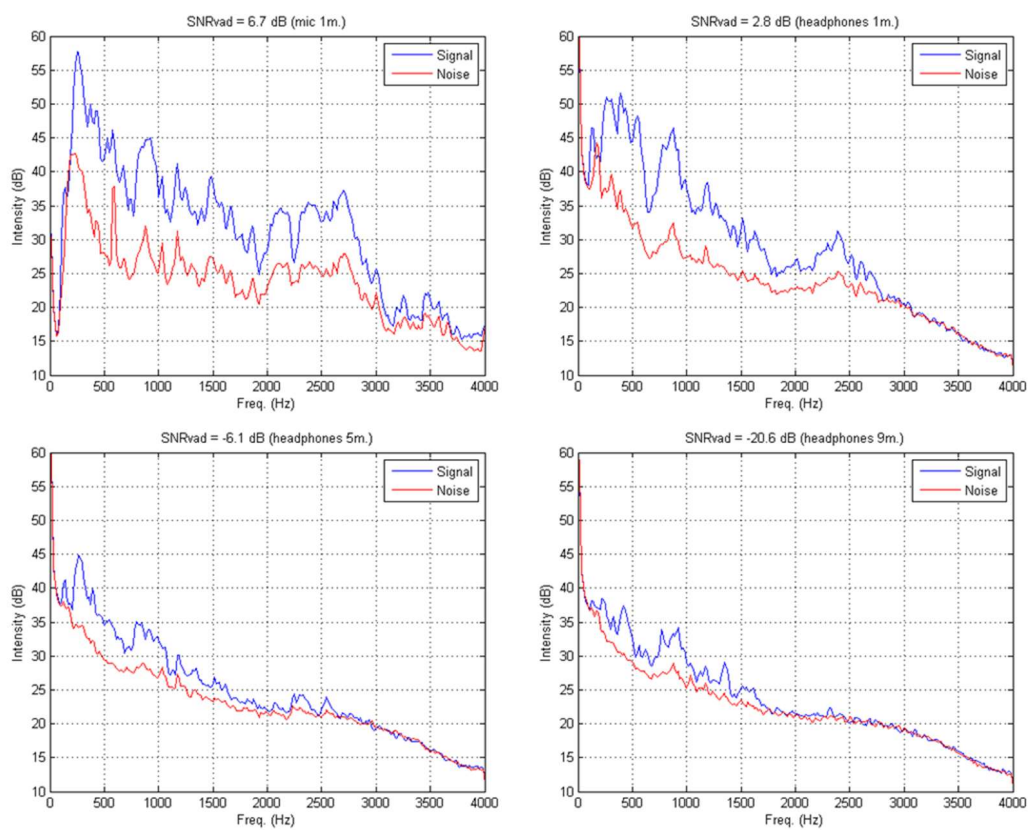

**Figure 4. Average signal and noise energy bands for microphone, one meter apart (top left); headphones, one meter apart (top right); headphones, five meters apart (bottom left); and headphones, nine meters apart (bottom right).**

Figure 4 displays average energy in voice-active regions (signal) compared to the average energy in voice-inactive (noise) regions. Note the sharp drop in the SNR in for the headphone channel setup for frequencies above around 1500 Hz in comparison with the microphone setup. High frequencies are further compromised as recording distance increases.

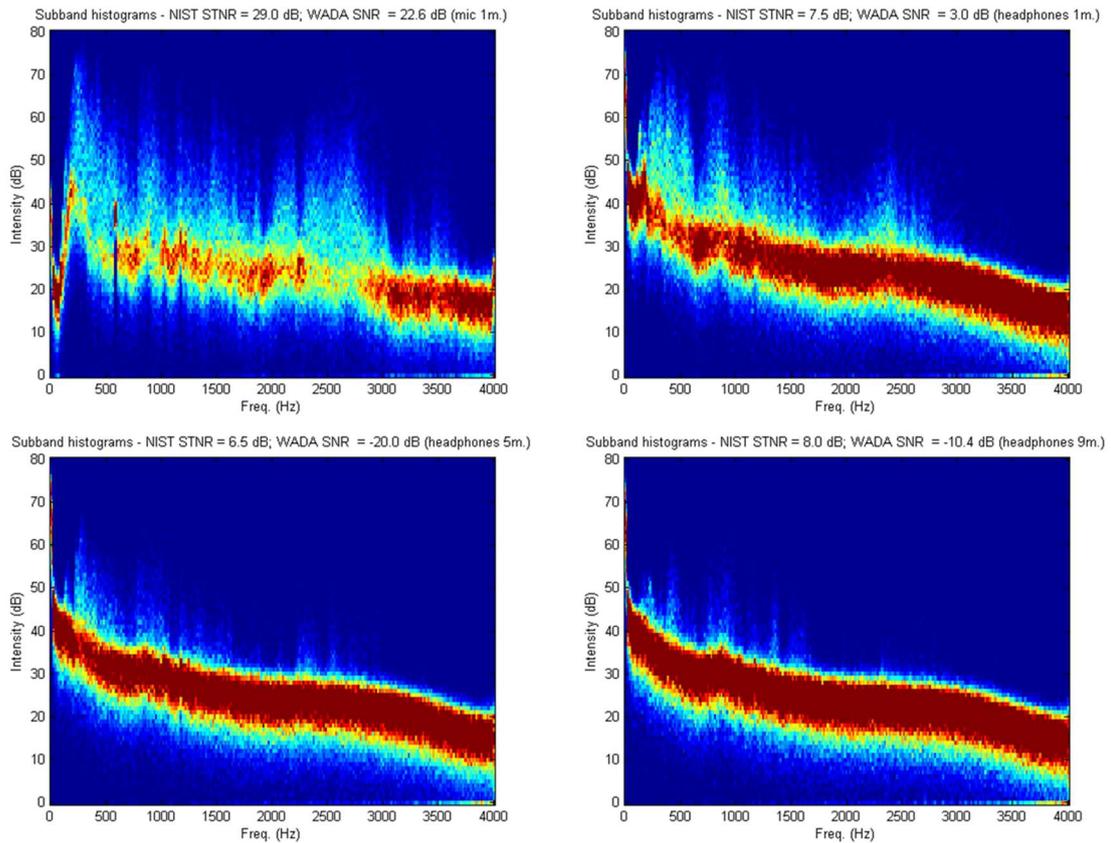

**Figure 5.** Energy histograms for microphone, one meter apart (top left); headphones, one meter apart (top right); headphones, five meters apart (bottom left); and headphones, nine meters apart (bottom right).

Figure 5 contains histograms for the dB energy levels for each frequency band. The histograms illustrate the lack of spectrum variability for the headphone spectra in comparison with that of the microphone. Note, as well, the relatively flat energy distribution for the headphones, especially at higher frequencies.

Figure 6 displays spectrogram views in which the darker regions correspond to signal energy in contrast to the brighter regions which correspond to noise. The spectrograms further emphasize the information loss found for the headphone recordings with respect to the microphone, particularly at higher frequencies.

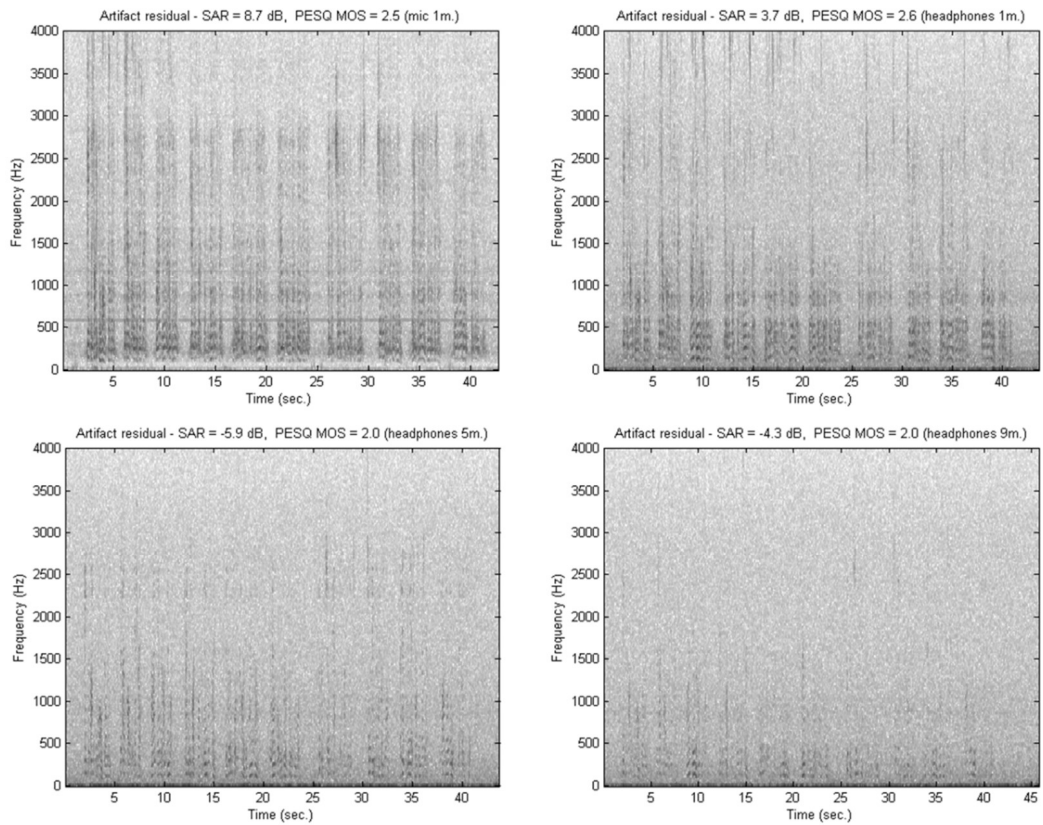

**Figure 6.** Spectrogram views for microphone, one meter apart (top left); headphones, one meter apart (top right); headphones, five meters apart (bottom left); and headphones, nine meters apart (bottom right).

The results portrayed in the figures and tables indicate that, among the speech quality measures utilized, the BSS_EVAL (SAR) and SNR VAD are those most correlated with reality. These measures consistently decrease with increasing distance, are far better for microphone recordings (versus headphone recordings), and decrease after AAC coding, as expected. The SAR index is of particular interest, since it is known to correlate, to a certain extent, with subjective ratings thus assessing speech intelligibility. Note, for instance, that the SAR index for a microphone positioned nine meters away from the speech source lies in between the indexes for headphones located three and five meters apart from the speech source. In addition, note that the codec's impact on degradation does not contribute to a substantial decrease in the speech quality. In general, our experiments suggest that reasonable an intelligible audio transmission can be achieved from a few meters using headphones as a microphone.

**Channel Capacity**

So far, we assessed the quality attained using headphones as microphones for speech transmission. In this chapter, we investigate the potential of using the headphone acquired acoustic waves to convey digital information, in terms of channel capacity. We focus on frequencies beyond the hearing range, which can be seen as secure and covert channels for transferring information between two computers.

Channel capacity (C) is a measure of the theoretical upper bound on the rate at which information can be transmitted (in bits per second) over a communication channel by means of signal S. Under the assumptions of additive interfering Gaussian noise (N) and available bandwidth (B (Hz)), the channel capacity can be calculated using the Shannon–Hartley theorem [17];

$$C = B \log_2 \left(1 + \frac{S}{N}\right)$$

The formula informs us that the higher the SNR and channel bandwidth, the higher the amount of information that can be conveyed. Note that for large SNRs (S/N >>1), $C \approx 0.33 \cdot B \cdot SNR\ (dB)$. Using this approximation, Figure 7 shows SNR values and respective channel capacity measured for different frequency ranges in our experimental setup, calculated as follows. Similarly to the previous experiments, pure sinusoidal tones were played from a source at different distances from the receiving computer and recorded via the headphones at a 44.1 kHz sample rate. The SNR was estimated for consecutive 100 Hz frequency bands up to 22 kHz, as the power ratio between the respective frequency tone and the average background noise over the bandwidth. SNR values were then used to estimate the channel capacity for each of the 100 Hz bandwidth sub-channels for different transmission distances, using the linear approximation described.

Our experiments suggest that headphones turned into microphone have significant potential for covert information transmission. In particular, considering that normal human hearing capabilities typically drop at frequencies over 10 kHz, and large inaudible spectrum regions are available for communication at reasonable rates.

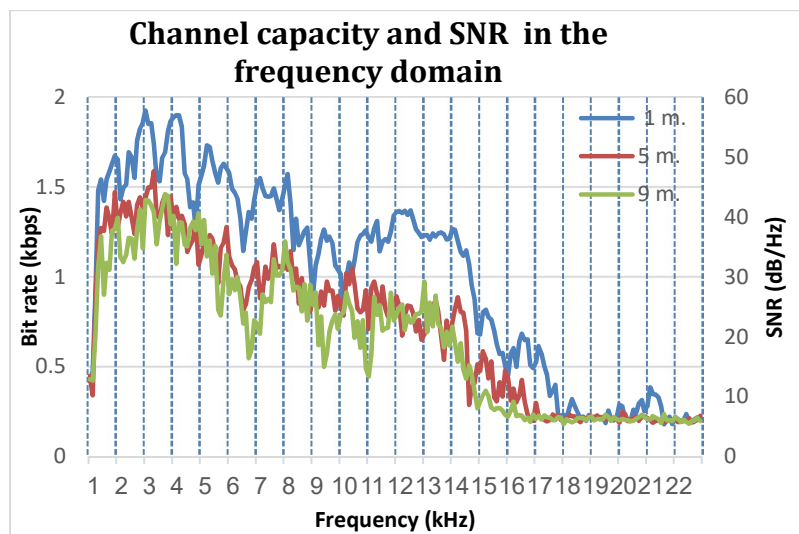

Figure 7. Channel capacity and SNR as a function of frequency.

**Improving SNR by Combining Channels**

The SNR can be reduced by averaging the outputs of two sampled channels. Assuming that the noise signals present at different channels are random and thus uncorrelated, summing the headphone/loudspeaker left and right channels would average out the noise, while enhancing the desired signals which are correlated. Theoretically, the uncorrelated noise sums as a root sum square, resulting in a 1.4 (square of two) increase, while perfectly correlated signals increase by a factor of 2. This difference yields a 3 dB increase in the SNR. Nevertheless, in practice, we did not succeed in improving the SNR of speech signals acquired through the headphones. We aligned and summed left and right headphone channels capturing speech from a distance of one meter, but the overall SNR gain obtained was a marginal 0.1 dB. We believe that due to the headphone channels proximity, there is a high level of correlation between the channels and the averaging process is not efficient.

## 5. Related Work

It is known that PC malware and mobile apps can use a microphone for spying purposes, and many types of spyware have been found in the wild [18] [19] [20] [21] [22] [23]. In 2015, Google has removed its listening software from the Chromium browser after complaints from privacy campaigners for potentially exposing private conversations [24]. More recently, Facebook was suspected of (and denied) using a mobile device's microphone to eavesdrop on conversations so it could better target ads [25]. In addition, there are currently many applications sold on the Internet that facilitate the use of microphones and cameras to gather information for surveillance and other purposes [26]. However, such mobile or desktop applications require either built-in or external microphones.

The general principle that an audio speaker is the exact inverse of an active microphone has been well known for years [3], as are the security concerns it raises [5] [27]. Lee et al. suggested turning the computer speaker into a microphone to establish covert communication between two loudspeakers at a limited distance of 10 centimeters [27]. Their work provides comprehensive measurements of different covert acoustic scenarios. However, most of the loudspeakers connected to PCs today have an integral amplifier which prevents passing any signal from output to input, and consequentially, beside [27]the threat of turning speakers into microphones in modern PCs for eavesdropping hasn't attracted much interest by security researchers so far.

## 6. Countermeasures

Countermeasures can be categorized into hardware and software countermeasures.

**Hardware Countermeasures**

In highly secure facilities it is common practice to forbid the use of any speakers, headphones, or earphones in order to create so-called audio gap separation [28]. Less restrictive policies prohibit the use of microphones but allow loudspeakers, however because speakers can be reversed and used as microphones, only active one way speakers are allowed. Such a policy was suggested by the NSTISSAM TEMPEST/2-95, RED/BLACK installation guide [29]. In this guide the protective measures state that "*Amplifiers should be considered for speakers in higher classified areas to provide reverse isolation to prevent audio from being heard in lesser classified areas.*" Some TEMPEST certified loudspeakers are shipped with amplifiers and one way fiber input [30]. Such a protective measure is not relevant to most modern headphones, which are primarily built without amplifiers. Other hardware countermeasures include white noise emitters and audio jammers which offer another type of solution aimed at ruining audio recordings by transmitting ambient sounds that interfere with eavesdroppers and don't allow them to accurately capture what is being said [31].

**Software Countermeasures**

Software countermeasures may include disabling the audio hardware in the UEFI/BIOS settings. This can prevent a malware from accessing the audio codec from the operating system. However, such a configuration eliminates the use of the audio hardware (e.g., for music playing, Skype chats, etc.), and hence may not be feasible in all scenarios. Another option is to use the HD audio kernel driver to prevent rejacking or to enforce a strict rejacking policy. For closed-source OSs (such as Microsoft Windows) such a driver must be developed and supported by audio codec vendors. In an improved approach, the kernel driver would prevent only out-to-in (speaker to mic) jack retasking, while enabling the use of other types of jack retasking. The kernel driver could also trigger an alert message when a microphone is being accessed, requesting explicit approval of such an operation from the user. In the same manner, anti-malware and intrusion detection systems can employ API monitoring to detect such unauthorized speaker-to-mic retasking operations and block them. A list of countermeasures, along with their pros and cons, is provided in Table 6.

Table 6. Countermeasures

| Countermeasure | Pros/Cons |
| --- | --- |
| Eliminate headphones/earphones/speakers | Pro: Hermetic protection |
|  | Con: Low usability |
| Using one way speakers | Pro: Hermetic protection |
|  | Con: Not relevant to headphones and earphones |
| BIOS/UEFI audio codec disable | Pro: Easy to deploy |
|  | Con: Low usability |
| Kernel driver policy enforcement | Pro: Easy to deploy |
|  | Con: Can be manipulated by software |
| Rejacking detection and alert | Pro: Easy to deploy |
|  | Con: Can be manipulated by software |
| White noise emitters/audio jammers | Con: Hard to deploy due to the environmental noise it generates |

## 7. Conclusion

Audio playing devices such as headphones and earphones (and certain types of loudspeakers) can be seen as microphones working in reverse mode: speakers convert electric signals into a sound waveform, while microphones transform sounds into electric signals. This physical fact alone may not pose a security threat, however modern PC and laptops motherboards include integrated audio codecs hardware which allow for modification of the audio jacks' functionality from output to input within software. In this paper we examine this issue in the context of cyber-security. We present SPEAKE(a)R a software that can render a PC, even once without microphones, into a eavesdropping device. We examine the technical properties of audio codec chips and explain why modern computers are vulnerable to this type of attack. We also present attack scenarios and evaluate the signal quality received by simple off the shelf headphones (with no microphone), when used as a microphones. Our experiments demonstrate that intelligible audio can be acquired through earphones and can then be transmitted distances up to several meters away. In addition, we showed that the same setup achieves channel capacity rates close to 1 Kbps in a wide range of frequencies.